\documentclass[aps,prl,preprint,superscriptaddress,floatfix,showpacs]{revtex4-1}

\usepackage{graphicx}
\usepackage{dcolumn}
\usepackage{bm}

\begin{document}


\title{
One-neutron knockout reaction of $\bm{^{17}\textrm{C}}$ on a hydrogen target at 70 MeV/nucleon}

\author{Y.~Satou}
\email[]{satou@snu.ac.kr}
\author{J.W.~Hwang}
\author{S.~Kim}
\author{K.~Tshoo}
\author{S.~Choi}
\affiliation{Department of Physics and Astronomy, 
Seoul National University, 599 Gwanak, Seoul 151-742, Republic of Korea}
\author{T.~Nakamura}
\author{Y.~Kondo}
\author{N.~Matsui}
\author{Y.~Hashimoto}
\author{T.~Nakabayashi}
\author{T.~Okumura}
\author{M.~Shinohara}
\affiliation{Department of Physics, 
Tokyo Institute of Technology, 2-12-1 Oh-Okayama, Meguro, Tokyo 152-8551, Japan}
\author{N.~Fukuda}
\author{T.~Sugimoto}
\author{H.~Otsu}
\author{Y.~Togano}
\author{T.~Motobayashi}
\author{H.~Sakurai}
\author{Y.~Yanagisawa}
\author{N.~Aoi}
\author{S.~Takeuchi}
\author{T.~Gomi}
\author{M.~Ishihara}
\affiliation{RIKEN Nishina Center, 2-1 Hirosawa, Wako, Saitama 351-0198, Japan}
\author{S.~Kawai}
\affiliation{Department of Physics, Rikkyo University, 
3 Nishi-Ikebukuro, Toshima, Tokyo 171-8501, Japan}
\author{H.J.~Ong}
\author{T.K.~Onishi}
\affiliation{Department of Physics, University of Tokyo, 
7-3-1 Hongo, Bunkyo, Tokyo 113-0033, Japan}
\author{S.~Shimoura}
\author{M.~Tamaki}
\affiliation{Center for Nuclear Study (CNS), University of Tokyo, 
2-1 Hirosawa, Wako, Saitama 351-0198, Japan}
\author{T.~Kobayashi}
\author{Y.~Matsuda}
\author{N.~Endo}
\author{M.~Kitayama}
\affiliation{Department of Physics, Tohoku University, 
Aoba, Sendai, Miyagi 980-8578, Japan}


\date{\today}

\begin{abstract}
First experimental evidence of the population 
of the first $2^-$ state in $^{16}\textrm{C}$ above the neutron threshold 
is obtained 
by neutron knockout from $^{17}\textrm{C}$ on a hydrogen target. 
The invariant mass method combined with in-beam $\gamma$-ray detection 
is used to locate the state at 5.45(1) MeV. 
Comparison of its populating cross section and parallel momentum distribution 
with a Glauber model calculation 
utilizing the shell-model spectroscopic factor 
confirms the core-neutron removal nature of this state. 
Additionally, 
a previously known unbound state at 6.11 MeV and 
a new state at 6.28(2) MeV 
are observed. 
The position of the first $2^-$ state, 
which belongs to a member of the lowest-lying $p$-$sd$ cross shell transition, 
is reasonably well described by the shell-model calculation 
using the WBT interaction. 
\end{abstract}


\maketitle

Much of our knowledge on quantum nature of atomic nuclei 
comes from studies of nuclear reactions 
in which an energetic beam of one nuclear species 
collides with a target made of another. 
Among various collision processes, 
the nucleon knockout reaction 
has become recognized as one of the most sensitive tools 
for spectroscopic studies, 
especially for nuclei away from the stability line, 
which include even those beyond the drip line. 
The knockout residue produced by removing a nucleon (or nucleons) 
from a fast moving beam particle, 
which impinges on a light target fixed in the laboratory, 
is observed in inverse kinematics 
by a detector placed in forward hemisphere efficiently. 
The removed nucleon(s) will be selected democratically from the valence space, 
allowing states with unique, often rarely accessible configurations 
to be populated in this process. 
The final state in the residue is identified 
by tagging de-excitation 
$\gamma$ rays~\cite{Navin08,Aumann00,Maddalena01} 
(see also references in Ref.~\cite{Gade08b}) 
and by observing decay neutrons and constructing the invariant 
mass~\cite{Simon99,Hoffman08,Hoffman09,Kondo10,Cao12,Lunderberg12,Aksyutina13,Kohley13}. 
For one-nucleon knockout case, 
the momentum spread of the residue reflects the Fermi motion of the nucleon 
suddenly removed, 
and is sensitive to the orbital angular momentum (the $l$ value) 
of the struck nucleon. 
Furthermore, 
the cross sections leading to individual final states 
relate to the occupancy of single-particle orbits, 
providing a link to details of the nuclear structure. 

The present study aims at exploring unbound states in $^{16}{\rm C}$ 
through an application of the one-neutron knockout technique 
to a $^{17}{\rm C}$ beam 
impinged on a proton target, 
for which simple reaction mechanisms are expected. 
Focus is placed in a search of lowest-lying cross shell transitions, 
the location of which reflects the shell gap between $p$ and $sd$ orbits. 
The neutron-rich carbon (C) isotopes 
have attracted attention as they often exhibit unique features: 
none of the odd mass C (heavier than $^{13}{\rm C}$) 
has the ground-state spin parity of $J^{\pi}_{\rm g.s.}$=5/2$^+$, 
the values which are expected from a naive shell model. 
There has been a debate about a reduced E2 transition strength 
(small proton collectivity) for the $2^+_1$ state 
in $^{16}{\rm C}$~\cite{Imai04,Elekes04,Ong08,Wiedeking08,Wuosmaa10,Petri12}. 
There is evidence for neutron halo formation for 
$^{15}{\rm C}$~\cite{Sauvan00}, 
$^{19}{\rm C}$~\cite{Nakamura99}, and 
$^{22}{\rm C}$~\cite{Tanaka10,Kobayashi12}. 
For some, if not all, of these features, 
nuclear deformation may play a key role, 
which occurs in this mass region 
due to near degeneracy of the $\nu d_{5/2}$-$\nu s_{1/2}$ orbits: 
neutrons in these orbits can gain energy by breaking spherical symmetry 
(the Jahn-Teller effect)~\cite{Hamamoto07}. 
The effect of nuclear deformation 
will further be signified by large quadrupole transition 
strengths~\cite{Elekes05,Ysatou08} 
and by a reduction of the major 
$p$-$sd$ shell gap~\cite{Talmi60,Suzuki97,Wiedeking05}. 
A recent $\beta$-delayed neutron 
emission study of $^{17}{\rm B}$~\cite{Ueno13} 
has reported low-lying negative parity states in $^{17}{\rm C}$, 
among which the lowest one was the $J^{\pi}$=1/2$^-$ state 
at the excitation energy of $E_x$=2.71(2) MeV. 
The energy of this state, 
reflecting the $p$-$sd$ shell gap, 
turned out to be 
lower than those of neighboring odd mass C isotopes:  
$E_x$=3.10 MeV for the $1/2^-_1$ state in $^{15}{\rm C}$ 
and $E_x$=3.09 MeV for the $1/2^+_1$ state 
in $^{13}{\rm C}$~\cite{Ajzenberg91}. 
This 
might indicate an onset of the $p$-$sd$ shell gap quenching 
towards heavier C isotopes. 
To examine this picture in more detail 
it is worthwhile to accumulate data 
on cross shell transitions in neighboring isotopes. 
This Letter reports on new relevant spectroscopic information 
on $^{16}{\rm C}$ in its unbound $E_x$ region. 
Besides, 
based on the parallel momentum distribution 
of the core fragment populated in a final state, 
it is demonstrated that the width of the distribution 
provides a good measure of the $l$ value (and thus the parity) 
of the state populated; 
for neutron knockout involving a proton target, 
this has previously been shown 
based on the transverse momentum distributions 
in the $^1{\rm H}(^{18}{\rm C},$$^{17}{\rm C}^*)$~\cite{Kondo09} 
and $^1{\rm H}(^{14}{\rm Be},$$^{13}{\rm Be}^*)$~\cite{Kondo10} reactions 
with the aid of elaborate reaction mechanism calculations. 

Despite relative proximity to stability, 
information on energy levels of $^{16}\textrm{C}$, 
particularly that above the neutron threshold 
(the neutron separation energy of $^{16}{\rm C}$ 
is $S_n$=4.250(4) MeV~\cite{Audi03}), 
has been limited. 
This is partially due to ineffectiveness of $\beta$ decay 
for this particular nucleus, 
as recognized by the absence of a parent nucleus 
($^{16}{\textrm B}$ is particle unstable). 
Early spectroscopic studies on $^{16}\textrm{C}$ 
utilized binary reactions involving transfers of neutrons. 
The $^{14}\textrm{C}(t,p)^{16}\textrm{C}$ two-neutron transfer 
studies~\cite{Fortune77,Balamuth77,Fortune78,Sercely78} 
have investigated levels below 7 MeV, 
including six bound states and an unbound state at 6.11 MeV. 
Since the ground state of $^{14}\textrm{C}$ 
is characterized by neutron $p$-shell closure, 
the states populated 
mostly involved configurations with two $sd$-shell neutrons, $(1s0d)^2$. 
The $^{13}\textrm{C}(^{12}\textrm{C},$$^9\textrm{C})^{16}\textrm{C}$ 
three-neutron transfer study~\cite{Bohlen03} 
has reported 14 more states up to $E_x$=17.4 MeV, 
including states with more complex configurations. 
Due to kinematical matching~\cite{Bohlen03} 
states with high angular momenta 
were favorably populated. 
Combining information 
from the recent $^{15}\textrm{C}(d,p)^{16}\textrm{C}$ reaction study 
using a radioactive $^{15}\textrm{C}$ beam~\cite{Wuosmaa10}, 
sound $J^{\pi}$ assignments 
have been available for bound states. 
For unbound states 
only the 8.92-MeV level 
has received a firm assignment of $5^-$~\cite{Bohlen03}. 
Two earlier one-neutron knockout studies on $^{17}{\rm C}$ using Be targets 
focused on transitions leading to bound final states 
in $^{16}{\rm C}$ by means of 
in-beam $\gamma$-ray spectroscopy~\cite{Maddalena01,Rodriguez-Tajes12}. 
They provided information 
not only on excited states of $^{16}{\rm C}$ 
but also on ground state properties of $^{17}{\rm C}$, 
e.g., the spin parity, 
$J^{\pi}_{\rm g.s.}(^{17}{\rm C})$=$3/2^+$, and 
no halo formation in spite of the remarkable low neutron separation energy 
of $S_n$=0.727(18) MeV~\cite{Audi03} 
due to a high angular momentum of $l$=2 for the valence neutron. 

The experiment was performed at the RIPS facility~\cite{Kubo92} of RIKEN. 
Details of the setup 
are provided in Refs.~\cite{Ysatou08,Tshoo12}, 
and a preliminary report of this work has been presented 
in Ref.~\cite{Hwang13}. 
The $^{17}\textrm{C}$ beam 
was produced from a 110-MeV/nucleon $^{22}\textrm{Ne}$ beam 
which impinged on a 6-mm-thick Be target. 
The typical $^{17}\textrm{C}$ beam intensity 
was 10.2 kcps with a momentum spread of $\Delta P/P$=$\pm 0.1\%$. 
The beam profile 
was monitored by a set of parallel-plate avalanche counters (PPACs) 
placed upstream of the experimental target. 
The target 
was pure liquid hydrogen~\cite{Ryuto05} contained in a cylindrical cell: 
3 cm in diameter, 120$\pm$2 mg/cm$^2$ in thickness, 
and having 6-$\mu$m-thick Havar foils for the entrance and exit windows. 
The average energy of $^{17}\textrm{C}$ at the middle of the target 
was 70 MeV/nucleon. 
The target was surrounded by a NaI(Tl) scintillator array 
used to detect $\gamma$ rays from charged fragments. 
The fragment was bent by a dipole magnet behind the target, 
and was detected by a plastic counter hodoscope 
placed downstream of the magnet. 
The $\Delta E$ and time-of-flight (TOF) information in the hodoscope 
was used to identify the $Z$ number of the fragment. 
The trajectory 
was reconstructed by a set of multi-wire drift chambers (MWDCs) 
before and after the magnet, 
which, together with TOF, 
gave mass identification. 
Neutrons were detected by two walls 
of plastic scintillator arrays placed 4.6 and 5.8 m downstream from the target. 
The neutron detection efficiency 
was $24.1\pm0.8\%$ for a threshold setting of 4 MeVee. 
The relative energy ($E_{\rm rel}$) of the final system 
was calculated from momentum vectors of the charged fragment and the neutron. 
In deducing the fragment vector, 
information on the impact point on target in transverse directions 
(determined by the upstream tracking detectors) 
was taken into account 
together with hit information within the MWDC placed behind the target. 
Neutron coincidence events 
were classified in terms of $E_{\rm rel}$ 
and the Fermi momentum of the struck neutron $k_3$. 
In the sudden approximation, 
the latter 
corresponds to the transferred momentum 
to the knockout residue ($^{16}{\rm C}$). 
The detector acceptance 
was evaluated by a Monte Carlo simulation 
as a function of $E_{\rm rel}$ and $k_3$. 

\begin{figure}[t]
\resizebox{0.90\columnwidth}{!}{%
\includegraphics[angle=-90]{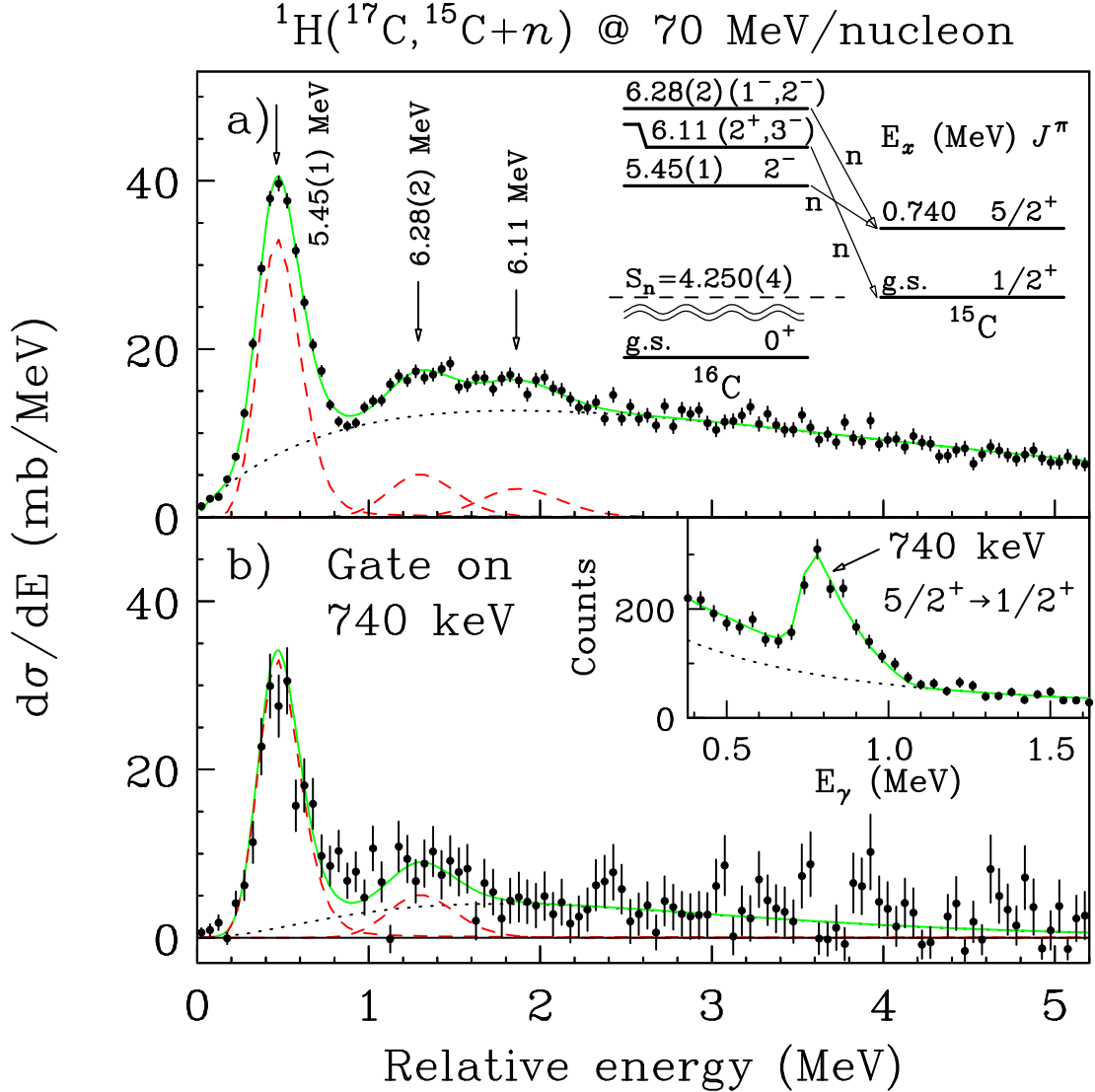}%
}
\caption{(Color online.) 
Relative energy spectra 
for the 
(a) $^1\textrm{H}(^{17}\textrm{C},$$^{15}\textrm{C}$+$n)$ and 
(b) $^1\textrm{H}(^{17}\textrm{C},$$^{15}\textrm{C}
(5/2^+; 0.74\ {\rm MeV})$+$n)$ 
one-neutron knockout reactions at 70 MeV/nucleon. 
Shown in the inset of panel (b) 
is the Doppler-corrected energy spectrum of $\gamma$ rays 
emitted from $^{15}{\rm C}$. 
Neutron coincidence is required for this spectrum. 
Green solid lines represent the results of the fit; 
dotted lines assumed background; 
red dashed lines extracted individual resonances. 
A decay scheme for states populated is shown in panel (a). 
\label{fig:spectrum_fit}}
\end{figure}

Figure~\ref{fig:spectrum_fit} 
shows relative energy dependence of cross sections for the 
(a) $^1\textrm{H}(^{17}\textrm{C},$$^{15}\textrm{C}$+$n)$ and 
(b) $^1\textrm{H}(^{17}\textrm{C},$$^{15}\textrm{C}
(5/2^+; 0.74\ {\rm MeV})$+$n)$ reactions at 70 MeV/nucleon. 
Background contributions measured by an empty target were subtracted. 
Error bars are statistical ones. 
Shown in the inset of Fig.~\ref{fig:spectrum_fit} (b) 
is the "Doppler-corrected" energy spectrum for $\gamma$ rays 
emitted from the decay product nucleus $^{15}{\rm C}$. 
A peak around $E_{\gamma}$=0.8 MeV 
arises from the decay of the first $5/2^+$ level at 0.74 MeV 
(the only bound excited state) 
to $^{15}{\rm C}_{\rm g.s.}$. 
The $5/2^+$ state 
is an isomeric state having a half-life of 2.61$\pm$0.07 ns~\cite{Ajzenberg91}. 
This 
long life time 
caused the emission point 
of the de-excitation $\gamma$ ray 
to be distributed along the path of the fast-moving decay product. 
The Doppler correction for the $\gamma$ ray energy 
was made by assuming that the decay occurs at 40.7 cm downstream of the target 
(an average decay point expected from the average beam energy 
and the known mean life for the isomeric state) 
in both data reduction and simulation by the \textsc{geant} code~\cite{GEANT}. 
The latter 
was done fully taking into account realistic geometry of the experiment. 
A higher energy tail for the photo-electric peak 
is due to this incomplete Doppler correction procedure. 
The simulated response, however, 
reproduced the data well 
as shown by the green solid line. 
The photo-peak efficiency over $E_{\gamma}$=0.60--1.12 MeV 
($\gamma$-ray window) 
was estimated to be 5.1(3)\% by the \textsc{geant} simulation. 
Figure~\ref{fig:spectrum_fit} (b) 
was obtained by gating on the $5/2^+$ $\gamma$ peak 
with the above window 
and by correcting for the detection efficiency. 
The background component 
was subtracted 
by assuming (i) 
that the background portion is the same 
as that in the $\gamma$-ray spectrum 
in the inset of Fig.~\ref{fig:spectrum_fit} (b): 
the portion of the area beneath the dotted line over the $\gamma$-ray window, 
which amounts to 46\%, 
and (ii) 
that the background shape is 
characterized by 
the inclusive spectrum in Fig.~\ref{fig:spectrum_fit} (a). 
A peak 
is clearly seen at $E_{\rm rel}$=0.46 MeV 
in Fig.~\ref{fig:spectrum_fit} (a). 
This is also evident in Fig.~\ref{fig:spectrum_fit} (b), 
indicating that 
this peak feeds the $5/2^+$ state in $^{15}{\rm C}$ after emitting a neutron. 
Besides, another resonance 
is visible at $E_{\rm rel}$$\approx$1.3 MeV 
in Fig.~\ref{fig:spectrum_fit} (b). 
These were observed for the first time. 

The relative energy spectrum of Fig.~\ref{fig:spectrum_fit} (a) 
was used in a fitting analysis 
to extract the parameters for the resonances: 
$E_{\rm rel}$ and the populating cross section $\sigma^{\rm exp}_{-1n}$. 
Their responses (dashed lines) 
were generated by a Monte Carlo simulation 
which takes into account the detector resolution, beam profile, 
Coulomb multiple scattering, 
and range difference inside the target. 
The relative energy resolution 
was estimated to scale as $\Delta E_{\rm rel}$=$0.17\sqrt{E_{\rm rel}}$ MeV 
(in rms). 
For the $E_{\rm rel}$=0.46-MeV state, 
a finite line width ${\mathit \Gamma}$ for an $l$=1 neutron emission 
was considered by adopting 
a single Breit-Wigner function~\cite{Lane-Thomas58,Ysatou08} 
(for other states, 
not as well isolated in the present data as this state, 
only the instrumental resolution was taken into account). 
In the analysis an excess strength 
was recognized at $E_{\rm rel}$$\approx$1.9 MeV. 
This may correspond to the nearest known state 
at $E_x$=6.11 MeV~\cite{Fortune77,Sercely78} 
if the decay product nucleus $^{15}{\rm C}$ is populated in the ground state 
(in Fig.~\ref{fig:spectrum_fit} (a) the response of this strength 
was created by assuming $E_{\rm rel}$=1.86 MeV: 
the difference between $E_x$=6.11 MeV and $S_n$ for $^{16}{\rm C}$). 
A similar fitting analysis using the $\gamma$-ray coincidence spectrum 
in Fig.~\ref{fig:spectrum_fit} (b), however, 
did not exclude the possibility that this component 
is absent from this spectrum. 
It is quoted that the upper limit on the fraction of $^{15}{\rm C}$ fragments 
from the decay of the $E_{\rm rel}$=1.86-MeV resonance, 
that were in the 0.74-MeV state 
is 20\% of the strength found in the spectrum 
in Fig.~\ref{fig:spectrum_fit} (a). 
The solid line in Fig.~\ref{fig:spectrum_fit} (a) 
shows the result of the fit to the total inclusive spectrum. 
The background (dotted line) coming from transitions to the continuum and 
from detecting neutrons 
not associated with the decay of excited states in $^{16}{\rm C}$, 
e.g., neutrons emitted from excited $^{17}{\rm C}$ nuclei 
that are created by inelastic scattering processes, 
was simulated by a function 
$a(E_{\rm rel})^b\exp(-cE_{\rm rel})$ 
with $a$, $b$, and $c$ free parameters. 
The results of the fit 
are summarized in Table~\ref{tbl:sigma_1n}. 
$E_x$ 
was calculated by $E_x$=$E_{\rm rel}$+$S_n$+$E^*$, 
with $E^*$ the excitation energy of $^{15}{\rm C}$. 
The errors include statistical ones 
and those due to the choice of the background shape 
(the latter, 
estimated by further assuming thermal emission of a neutron ($b$=1/2) 
and neutron evaporation ($b$=1) for the background shape, 
turned out to be a dominant source of the error: 
they were 80, 90, and 70\% of the errors quoted in Table~\ref{tbl:sigma_1n} 
for $E_{\rm rel}$, ${\mathit \Gamma}$, and $\sigma^{\rm exp}_{-1n}$, respectively, 
for the $E_{\rm rel}$=0.46-MeV state, 
while they 
were 90, 20, and 70\% in the (upper bounds of) errors 
of $\sigma^{\rm exp}_{-1n}$ for the $E_{\rm rel}$=1.86-MeV state 
and of $E_{\rm rel}$ and $\sigma^{\rm exp}_{-1n}$ 
for the $E_{\rm rel}$=1.29-MeV state, respectively). 
For $\sigma^{\rm exp}_{-1n}$, 
uncertainties originating from the target thickness and 
neutron detection efficiency 
are included. 
The same fit was repeated 
for the $\gamma$-ray coincidence spectrum of Fig.~\ref{fig:spectrum_fit} (b) 
using the responses 
for the $E_{\rm rel}$=0.46 and 1.29-MeV states, 
obtained from the fit to the inclusive spectrum 
in Fig.~\ref{fig:spectrum_fit} (a). 
To quantify the character of the latter state, 
as a state built on the $^{15}{\rm C}^*$(0.74 MeV) excited core, 
the fit was repeated by changing the strength from the original one. 
By finding the fractional value at which the $\chi^2$ of the fit 
alters from the minimum by one unit, 
a lower limit of 32\% was deduced. 
The extraction of the $l$ value of the knocked-out neutron 
from a differential quantity 
for the $E_{\rm rel}$=0.46-MeV state 
is explained later. 


To allow discussion in terms of nuclear structure, 
reaction model calculations based on the Glauber 
approximation~\cite{Hansen03} 
were performed. 
The one-neutron removal cross section $\sigma^{\rm th}_{-1n}$ 
is expressed for a given final state with $J^{\pi}$ as 
\begin{equation}
\sigma^{\rm th}_{-1n}=\sum_{nlj}\left( \frac{A}{A-1}\right)^{N}
C^2S(J^{\pi},nlj)
\sigma_{\rm sp}(nlj,S_n^{\rm eff}), 
\end{equation}
where $A$ is the projectile mass, $N$ the major oscillator quantum number, 
$C^2S$ the spectroscopic factor, 
and $\sigma_{\rm sp}$ the single-particle cross section. 
The quantum numbers of the removed neutron are denoted by $nlj$. 
$S_n^{\rm eff}$ is the effective separation energy given by 
the sum of $S_n$ of the projectile 
and $E_x$ of the residue. 
$\sigma_{\rm sp}$ 
was calculated by the code \textsc{csc\_gm}~\cite{Abu-Ibrahim03} 
taking into account 
both stripping and diffractive processes 
(effective nature of the nucleon-nucleon ($NN$) profile function used 
resulted in small non-zero stripping cross sections)~\cite{Hansen03}. 
The elastic $S$ matrix 
for the collision 
of the residue (core) with the proton target 
was calculated by folding the finite-range Gaussian $NN$ 
profile function~\cite{Abu-Ibrahim08} 
with the point proton and neutron densities of the core 
obtained from 
the Hartree-Fock (HF) calculation using the SkX interaction~\cite{Brown98}. 
The $S$ matrix for describing the scattering of 
the valence neutron with the target proton 
was given by $S(b)$=1$-{\mathit \Gamma}_{pn}(b)$, 
here $b$ is the impact parameter of the colliding nucleons, 
and ${\mathit \Gamma}_{pn}$ 
the profile function for proton-neutron scattering. 
The parameters chosen for the profile function 
are those 
describing the $NN$ total and elastic cross sections 
consistently~\cite{Abu-Ibrahim08}.
The neutron-residue relative motion 
was calculated in a Woods-Saxon potential. 
The depth was adjusted 
so as to reproduce $S_n^{\rm eff}$, 
for a diffuseness $a_0$=$0.7$ fm and a reduced radius $r_0$ specifically chosen 
to be consistent with the HF calculation~\cite{Terry04,Gade08}: 
$r_0$ generates a single-particle 
wave function with a rms neutron-core separation 
of $r_{\rm sp}$=$[A/(A-1)]^{1/2}r_{\rm HF}$ at the HF-predicted binding energy, 
where $r_{\rm HF}$ is the HF rms radius of each orbit. 
The spin-orbit potential 
had the same $a_0$ and $r_0$ as the central one with a strength of $-12$ MeV 
in the notation of Ref.~\cite{Bertulani06}. 
The HF radius for the $p_{1/2}$ ($p_{3/2}$) orbit of $^{17}\textrm{C}$, 
for example, 
is 
2.966 (2.779) fm; 
this translates into $r_{\rm sp}$=3.057 (2.865) fm, 
which is reproduced by taking $r_0$=$1.263$ (1.234) fm. 
The $C^2S$ values 
were obtained by the shell-model code \textsc{nushell}~\cite{nushell} 
using the WBT interaction~\cite{Warburton92} in the $spsdpf$ model space. 
The calculated results 
for relevant states 
are given in Table~\ref{tbl:sigma_1n}. 
$\sigma^{\rm th}_{-1n}$ 
includes contributions from both stripping ($\sigma_{\rm str}$) and 
diffractive ($\sigma_{\rm diff}$) mechanisms. 
Due to inert nature of the proton, 
the latter dominates the knockout processes. 

\begin{table}[t]
\caption{States populated by 
the $^1{\rm H}$($^{17}{\rm C}$,$^{16}{\rm C}$) reaction. 
Theoretical cross sections were obtained by using the Glauber-model code 
\textsc{csc\_gm}~\cite{Abu-Ibrahim03} and the shell-model spectroscopic factors 
calculated with the WBT interaction~\cite{Warburton92}. 
Calculations 
used $S_n^{\rm eff}$ involving experimental $E_x$ values. 
\label{tbl:sigma_1n}} 
\begin{ruledtabular}
\begin{tabular}{ccccccccccc}
\multicolumn{4}{l}{Experiment}  & &
\multicolumn{3}{l}{Theory} \\
\cline{1-5} 
\cline{6-10}
$E_{\rm rel}$   & 
$E_x$   & 
${\mathit \Gamma}$ & 
$l$ &
$\sigma^{\rm exp}_{-1n}$ & &
$E_x$   & 
$\sigma_{\rm str}$  & 
$\sigma_{\rm diff}$  & 
$\sigma^{\rm th}_{-1n}$  & 
$J^{\pi}$     \\ 
(MeV)  & (MeV) & (MeV) & ($\hbar$) & (mb)  &  & 
(MeV)  & (mb)  & (mb)  & (mb)  & 
    \\ \hline
0.463(3)\protect\footnotemark[1] & 5.45(1) & 0.03(1) & 1 & 10.6(6) & & 
5.57 & 1.38 & 14.23 & 15.61 & \multicolumn{1}{c}{$2^-_1$} \\
1.86\protect\footnotemark[2]  & 
6.11 & 
--- & --- & 
$2.0^{+0.4}_{-0.8}$ 
& &  
5.75  & 0.05 & 0.53 &  0.58 & ($3^-_1$) \\
      &                              &     &     &           & &  
7.60  & 0.09 & 1.37 &  1.46 & ($2^+_3$) \\
      &                              &     &     &           & &  
8.81  & 0.04 & 0.31 &  0.35 & ($4^+_2$) \\
1.29(2)\protect\footnotemark[1]  & 6.28(2) & --- & --- & 
$2.5^{+0.2}_{-1.9}$
& &  
6.55  &  0.61  & 5.43  & 6.04 & ($1^-_2$) \\
                                 &         &     &     &          & &  
6.63  &  0.28  & 2.57  & 2.85 & ($2^-_2$)
\end{tabular}
\footnotetext[1]{Observed in coincidence with the 0.74-MeV 
$\gamma$ ray from $^{15}{\rm C}$. }
\footnotetext[2]{Derived from the energy $E_x$=6.11 MeV 
in Ref.~\cite{Fortune77} 
by assuming the $^{15}{\rm C}$ core is in the ground state. }
\end{ruledtabular}
\end{table}

The state observed at $E_x$=5.45 MeV 
was found to be well explained by the $2^-_1$ shell-model state 
in both position and cross section, 
making an assignment of $2^-$ appropriate. 
The $2^-_1$ state 
exhibited the highest cross section of unidentified shell-model states 
in the energy region of interest. 
The summed $\sigma^{\rm th}_{-1n}$ for predicted $2^-$ and $1^-$ states 
below 8 MeV, 
where major strengths are concentrated, 
are 20.9 
(15.5, 2.8, and 2.6 mb at $E_x$=5.57, 6.63, and 7.23 MeV, respectively) 
and 11.5 mb (see below for composition), respectively. 
Their ratio is near to the statistical ratio of 5:3 
expected for a doublet with spins $J$=2 and 1, 
allowing an interpretation that the $2^-$ and $1^-$ states 
are formed by coupling a hole in the $\nu p_{1/2}$ orbit 
to three neutrons with $J$=3/2 in the $sd$ orbits 
(note that $J^{\pi}_{\rm g.s.}(^{17}{\rm C})$=$3/2^+$). 
The predicted $1^-$ strength is distributed 
among states at $E_x$=5.79, 6.55, and 6.98 MeV 
with $\sigma^{\rm th}_{-1n}$=0.6, 6.0, and 4.9 mb, respectively. 
The fragmentation of the strength 
and the general trend in the Glauber model 
to overestimate the cross section~\cite{Gade08b,Gade08} 
would exclude 
an assignment of $1^-$ for the 5.45-MeV state 
with $\sigma^{\rm exp}_{-1n}$=10.6(6) mb. 

\begin{figure}[t]
\resizebox{0.90\columnwidth}{!}{%
\includegraphics[angle=0]{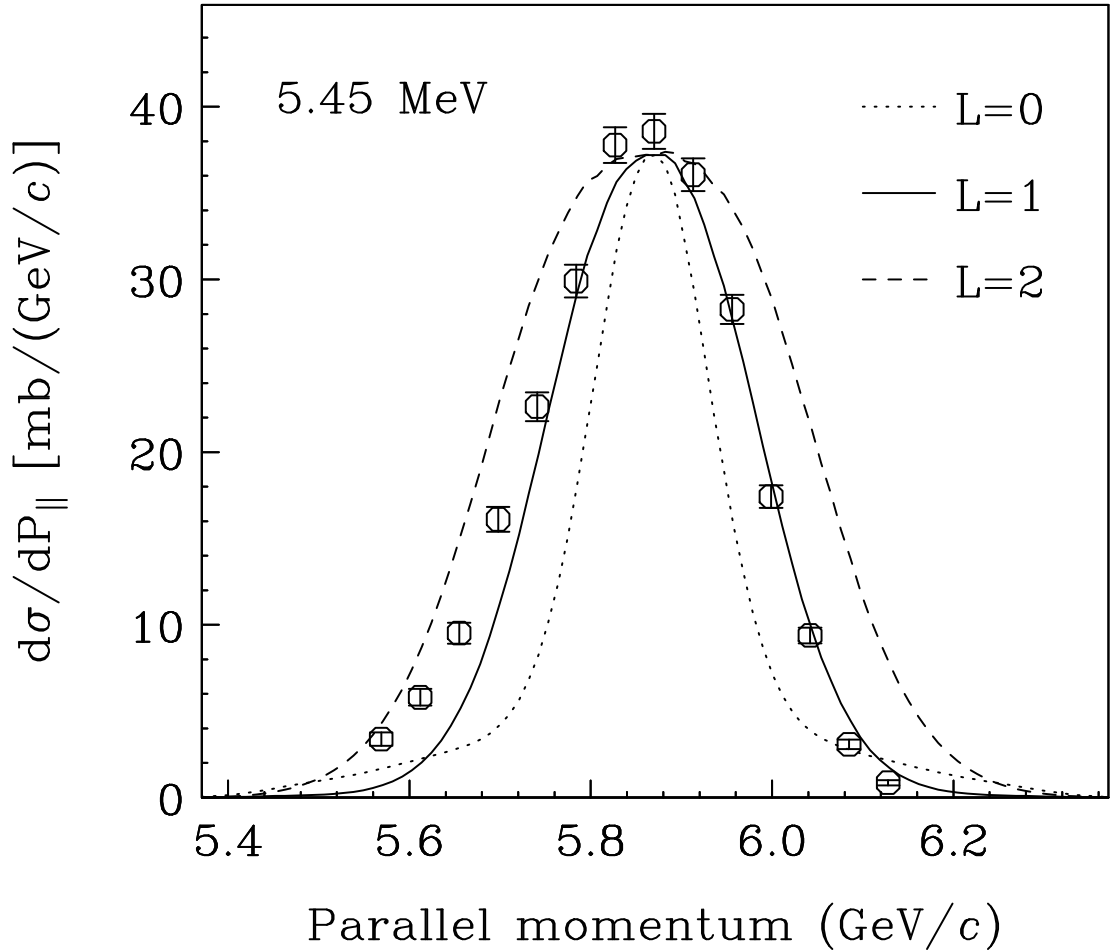}%
}
\caption{Laboratory $p_{||}$ distribution of $^{16}{\rm C}$ populated 
in the 5.45-MeV state after knockout from $^{17}{\rm C}$ (open circles). 
The dotted, solid, and dashed lines 
are the Fourier transform of single-particle wave functions 
of orbitals with $l$=0, 1, and 2, respectively. 
\label{fig:pl}}
\end{figure}

Figure~\ref{fig:pl} 
shows the laboratory parallel momentum ($p_{||}$) distribution 
leading to the 5.45-MeV state. 
This was obtained by subdividing, in terms of $p_{||}$, 
the inclusive spectrum 
and repeating the fitting procedure described above. 
The errors are statistical ones. 
Also plotted in Fig.~\ref{fig:pl} 
are the $p_{||}$ distributions calculated with \textsc{csc\_gm} 
for varying $l$ values. 
An experimental resolution of 43(1) MeV/$c$ in rms is convoluted. 
Factors relevant to stripping mechanisms are dropped, 
and the curves 
represent the Fourier transform of the single-particle wave functions. 
The full width at half maximum (FWHM) of the experimental distribution 
for the 5.45-MeV state 
was determined by a fit using a Gaussian 
to be 210(11) MeV/$c$ after unfolding the resolution. 
In the fit, 
a low-energy tail ($p_{||} <$ 5.72 GeV/$c$), 
which often suffers from higher-order effects~\cite{Tostevin02}, 
was 
eliminated. 
The fit curve (not shown) is similar to the $l$=1 curve (solid line). 
The width 
agrees well with 233 MeV/$c$ FWHM 
calculated 
for $p$-wave knockout, 
whereas for $s$- (dotted line) and $d$-wave (dashed line) knockout, 
widths of 121 and 377 MeV/$c$ FWHM were respectively predicted, 
incompatible with the measurement. 
This observation 
agrees to the expected character 
of the 5.45-MeV, $2^-$ state as having a neutron hole in the $p$ orbit, 
illustrating the robust feature of the $p_{||}$ distribution 
as an $l$ identifier. 

The large populating cross sections observed for the 6.11-MeV state 
in the $^{14}{\rm C}(t,p)^{16}{\rm C}$ reactions~\cite{Fortune77,Sercely78} 
have suggested that this state is either of the natural parity 
$2^+$, $3^-$, or $4^+$ states. 
The knockout cross sections calculated for the relevant 
$2_3^+$, $3_1^-$, and $4_2^+$ shell-model states, 
together with their shell-model energies, 
are compared to the data in Table~\ref{tbl:sigma_1n}. 
The present data turned out not to provide a strong constraint 
on the $J^{\pi}$ values for this state, 
although in terms of comparisons in both $E_x$ and $\sigma_{-1n}$ 
they seem to prefer the assignment of $2^+$ or $3^-$. 
The 6.28-MeV state exhibited the same decay pattern as the strongest 
5.45-MeV $2^-$ state 
with a sizable cross section. 
The $1^-_2$ and $2^-_2$ states 
predicted at 6.55 and 6.63 MeV, respectively, 
had large populating cross sections and are candidates for this state. 


\begin{figure}[t]
\resizebox{0.90\columnwidth}{!}{%
\includegraphics[angle=0]{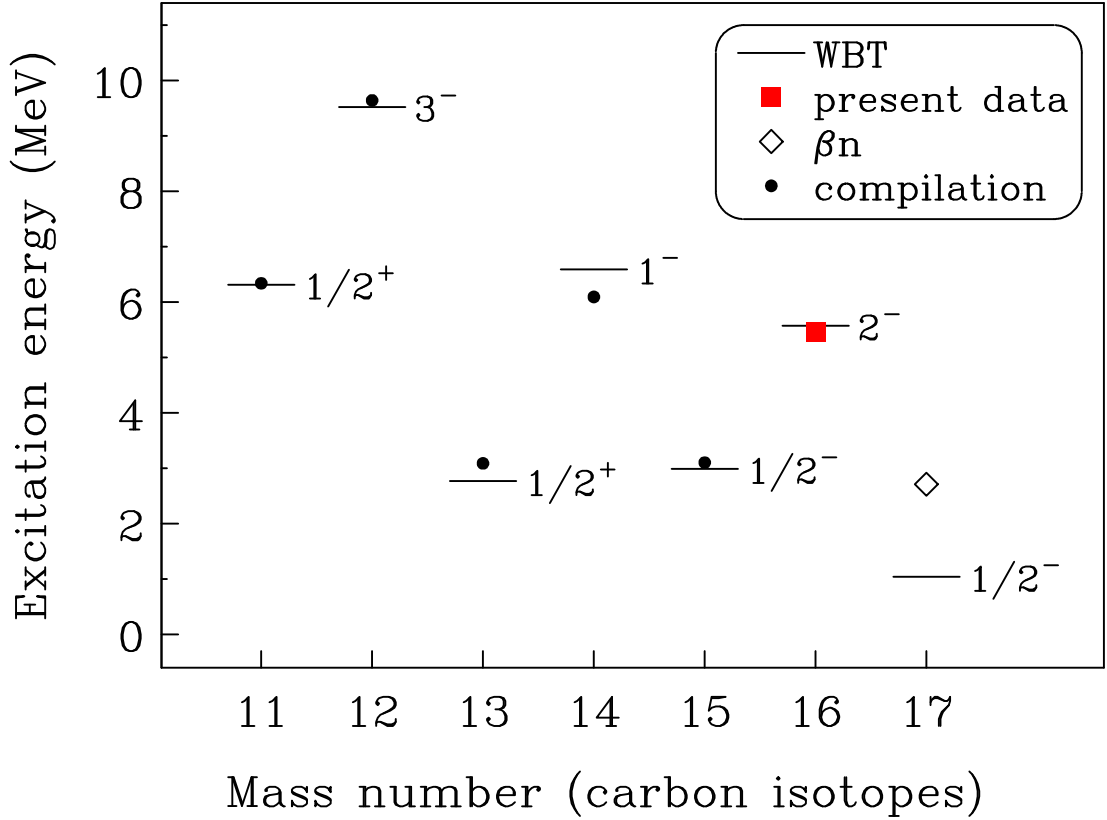}%
}
\caption{(Color online.) 
The migration of energies of (known) lowest-lying states in C isotopes, 
whose parities are opposite to those of their respective ground states, 
in comparison to shell-model values 
obtained by using the WBT interaction~\cite{Warburton92}. 
Data for $^{11-15}{\rm C}$ (filled circles) 
are from Refs.~\cite{Ajzenberg91,Ajzenberg90}. 
The data point for $^{16}{\rm C}$ (red filled square) 
is from the present study, 
while that for $^{17}{\rm C}$ (open diamond) is from Ref.~\cite{Ueno13}. 
The shell-model calculations 
were performed within the 2$\hbar \omega$ and 0$\hbar \omega$ bases 
(for both positive and negative parity states) 
for $^{11-15}{\rm C}$ and $^{16,17}{\rm C}$, respectively. 
\label{fig:wbt_cross_shell_states}}
\end{figure}

The presently observed $2^-$, 5.45-MeV state in $^{16}{\rm C}$ 
belongs to a member of the lowest-lying states 
having 
an opposite parity to the ground state. 
The location of such states provides a measure 
of the $p$-$sd$ shell gap and 
it is well explained by the shell model using the WBT interaction 
across the C isotopes, $^{11-15}{\rm C}$. 
To illustrate the latter, 
their energies are compared 
to the shell-model values 
in Fig.~\ref{fig:wbt_cross_shell_states}. 
In a recent study of $\beta$-delayed neutron emission 
of $^{17}{\rm B}$~\cite{Ueno13}, 
three low-lying negative parity states 
were newly identified in just one-neutron heavier nucleus $^{17}{\rm C}$. 
The WBT interaction turned out to fail 
in predicting their location by about 1 MeV 
(theory predicts lower values, see also Fig.~\ref{fig:wbt_cross_shell_states}), 
and several possible mechanisms, 
such as reduction in pairing energy for neutrons in the $sd$ orbits, 
were discussed. 
The present study adds a case 
in which the shell model with the WBT interaction 
predicts the location of the lowest-lying cross shell transition properly 
(see also Fig.~\ref{fig:wbt_cross_shell_states}), 
showing 
that this interaction 
describes the $p$-$sd$ shell gap in $^{16}{\rm C}$ adequately. 
To pin down the source of the discrepancy 
between theory and experiment 
on the position of the cross shell transition in $^{17}{\rm C}$, 
as discussed in Ref.~\cite{Ueno13}, 
and to better understand the dynamical evolution of single-particle orbits 
and relevant residual interactions 
away from stability, 
further spectroscopic studies on such states 
in heavier C as well as neighboring isotopes are of help. 

In summary, 
one-neutron knockout from $^{17}{\rm C}$ on a proton target 
was exploited in populating two new states at 5.45(1) and 6.28(2) MeV, 
and a previously known state at 6.11 MeV in $^{16}{\rm C}$. 
The energy spectrum 
was constructed utilizing the invariant mass method 
involving a decay neutron and a $^{15}{\rm C}$ fragment. 
De-excitation $\gamma$ rays from the latter 
were measured to correctly locate the resonances. 
For the 5.45-MeV state, 
an attempt was made to deduce the orbital angular momentum 
of the knocked-out neutron 
from the parallel momentum distribution 
associated with the unbound knockout residue. 
This, together with a comparison in terms of the measured and calculated 
knockout cross sections, 
has led to a spin-parity assignment of $2^-$ for this state. 
Possible spins and parities have been suggested for the other states, 
bringing about an advanced understanding 
of the level scheme of $^{16}{\rm C}$. 
The energy of the first $2^-$ state was adequately reproduced by the 
standard shell-model calculation using the WBT interaction 
without invoking modifications to the residual interaction. 

\begin{acknowledgments}
This work was supported in part by 
the Grant-in-Aid for Scientific Research (15740145) of MEXT Japan 
and the NRF grant (R32-2008-000-10155-0 (WCU), 2010-0027136) of MEST Korea. 
\end{acknowledgments}

\end{document}